\documentstyle[12pt]{article}          % usual 12pt

\catcode`\@=11
\@addtoreset{equation}{section}

	\newdimen\eqskip
	\newdimen\txtskip
	\baselineskip=\txtskip
\begin{document}
\def    \s2t           {\sin^2 2 \theta_0~}
\def    \dmsq          {\Delta m^2_0~}

%  ABSTRACT

\begin{flushright}
Fermilab-Conf-93/056-T\\
\today
\end{flushright}

\vspace*{0.5in}
\begin{center}
       	{ \Large  \bf \sc
OVERVIEW OF ACCELERATOR LONG BASELINE \\[0.1in]
NEUTRINO OSCILLATION EXPERIMENTS
	}
\footnote{Invited talk at the ``Moriond Workshop on Neutrino
Physics'', January 30 to February 6, 1993 at Villars, Switzerland.}

\vskip 0.8in

{\it STEPHEN PARKE }\\[0.1in]
{\small Fermi National Accelerator Laboratory,\\
P.O. Box 500, Batavia, IL 60510, U.S.A.} \\[0.2in]
\end{center}

\vskip 1.5in
\begin{abstract}
There is renewed interest in performing
a long baseline neutrino oscillation experiment using
accelerator neutrinos because of a discrepancy between
the measured  and the predicted  values
of the ratio of electron  to muon neutrinos
produced in the upper atmosphere
by cosmic rays.
The approximate range in oscillation parameter
space indicated by
the Kamiokande atmospheric neutrino
results and confirmed by IMB is
bounded by $ 10^{-3}~ eV^2  <  \dmsq  <  10^{-1}~ eV^2 $
and $\s2t  > 0.4~$.
To reach such small $\dmsq$, using
an accelerator as the source of neutrinos where the  energy is
typically 1 GeV or greater,
requires baselines
in the range of 10~km to 1000~km.
In this talk I will give an overview of  the most likely
possibilities for such a long baseline accelerator
neutrino oscillation experiment.

\end{abstract}
%--------------------------------------------
%       DEFINITIONS

% Text shorthands
\def	\epslon		{\mbox{$P_{min}$}}
\def    \plot0          {\mbox{$(\sin^2 2 \theta_0,~\Delta m^2_0)~~$}}
\def    \plotN          {\mbox{$(\sin^2 2\theta_N,~\Delta m^2_N)~~$}}
\def    \mue          {\mbox{$\nu_{\mu} \leftrightarrow \nu_e $}}
\def    \mutau        {\mbox{$\nu_{\mu} \leftrightarrow \nu_{\tau}$}}
\def    \figfs        {1}
\def    \figbn        {2}
\def    \figci        {3}
\def    \figcg        {4}
\def    \figkk        {5}

\newpage

\section {Introduction}
The recent indications of a deficit in the $\nu_{\mu}$ flux
of atmospheric neutrinos and the long-standing solar
neutrino problem have motivated new searches for neutrino
oscillations with small neutrino $\Delta m^2$.
The  results of the Kamiokande collaboration on
atmospheric neutrinos suggest oscillation parameters in the range
bounded by $ 10^{-3}~ eV^2  <  \dmsq  <  10^{-1}~ eV^2 $
and $\s2t  > 0.4~$. This result has renewed interest
in using accelerator neutrinos to perform a long
baseline neutrino oscillation experiment. Many
possibilities have been discussed; it is almost like
picking an accelerator from column A and a detector from
column B and there you have a possible experiment!
In this overview I will brief review the range of parameter space
a neutrino oscillation experiment can explore
and then discuss  the most likely possiblities.

The transition probability of producing one flavor of
neutrino $\nu_a$, of energy $E$, at the source,
letting the neutrino propagate to
the detector, a distance $L$ away, and then detecting the
neutrino as a different flavor $\nu_b$, is
(for a derivation see, for example, ref \cite{bp})
\begin{equation}
\label{vprob}
        {\cal P}_{ab} \;=\; \sin^2 2 \theta_0 \;\; \sin^2 \left( 1.27{
{\Delta m^2_0 \; L} \over {E}} \right) \end{equation}
where $\Delta m^2_0 ,~E$ and $L$ are measured in
${\rm eV}^2$, GeV, and km respectively.
$\Delta m^2_0$ is the difference of the squares of the masses
 for the two
neutrino mass eigenstates and $\theta_0$ is the mixing angle
relating these mass eigenstates to the flavor eigenstates.
The experiments that measure this probability
either measure a finite value for ${\cal P}_{ab}$ or assign
a limit ${\cal P}_{ab} < \epslon $; the value of
$\epslon $, the energy spectrum of detected neutrinos
and the source--detector distance then
 define a region in the {\mbox{$(\sin^2 2 \theta_0,~\Delta m^2_0)~~$}}plane
for each experiment.
This $\epslon $
is the minimum measurable oscillation probability for the experiment
in a given analysis mode.
The size of  $\epslon $, or the limit in our ability
to measure ${\cal P}_{ab}$, arises from many sources;
statistical uncertainties,
the contamination of the beam with  other neutrino species,
the fractional uncertainty in the neutrino flux calculations,
the knowledge of the experimental acceptance
for the different neutrino species, the backgrounds to the $\nu_b$
signal and many other systematic uncertainties.

For large $\Delta m^2_0$ ($\gg E/L$) an experiment can explore  any
\begin{equation}
\label{vssqlt} {\sin^2 2 \theta_0} \geq 2 ~\epslon.
\end{equation}
For $ {\sin^2 2 \theta_0} \;=\; 1$ the limit on the mass difference squared is
\begin{equation}
\label{vmlt}
        \Delta m^2_0 \geq { \sqrt{\epslon} \over 1.27} { E \over L },
\end{equation}
assuming $\epslon << 1$.
For smaller $ {\sin^2 2 \theta_0} \,$ a good
{\it approximation} to the probability contour is  a straight
line
with slope $-1/2$ in a log-log plot in the
{\mbox{$(\sin^2 2 \theta_0,~\Delta m^2_0)~~$}plane until this line
intersects the vertical line from Eq.~\ref{vssqlt}.

Therefore, for the range indicated
by the atmospheric neutrino data, an
interesting experiment for this purpose
will need to have a $ \epslon \approx 10\% $ or better
and $ E/L $ at least one
but preferable two orders of magnitude
smaller than 0.4 GeV/km.

\section{Fermilab - Soudan 2}

The new Main Injector at Fermilab will be a 120 GeV proton accelerator
that can deliver $2 \times 10^{20}$ protons
on target (POT) in a $10^7$sec-year.
If a new neutrino beamline is constructed at Fermilab
for both a short baseline experiment, P803 \cite{FNALb},
as well as for a long baseline experiment, using
this accelerator as a source of protons, then
the average energy of muon neutrinos produced would be approximately
10 GeV. The contamination from muon  anti-neutrinos  and
electron neutrinos will be at the percent level.
Initially there were three proposals for the long baseline
 detector: IMB, Soudan 2
and DUMAND. Since that time, IMB has ceased to exist and the engineering
 and
environmental issues of sending a beam 30 degrees below
the horizontal to DUMAND, make it prohibitively expensive.

Soudan 2 is 710 km from
Fermilab in a direction 3.2 degrees
below the horizontal \cite{FNALa}.
This detector is a modular fine-grain tracking
calorimeter with a mass of 1 kTon
(with a possible upgrade  to 5 kTon)
surrounded on all sides by a two-layer
active shield of proportional tubes.
The analysis
would be based on the ratio of events that appear in the far detector to
be of neutral current type to that of charged current type
 compared to the same ratio in a near
``identical'' detector.
The exclusion plots are shown in
Fig.\figfs.
The difference in the limits between the two oscillation modes
comes from the difference in the
$\nu_{\tau}$ charged current cross section compared to $\nu_e$ or
$\nu_{\mu}$.

\begin{figure}[hb]
\vspace{5.75in}
\label{fs}
\caption{Fermilab/Soudan 2 90\% CL limits:
A and B are for a 4 year run and a 5
kTon fiducial volume detector (14k events)
for  $\nu_{\mu} \leftrightarrow \nu_e $
and  $\nu_{\mu} \leftrightarrow \nu_{\tau}$ respectively.
C and D are for a 9 month run using only the current 0.9 kton (680 events)
for \mue~ and \mutau~ respectively.}
\end{figure}

\newpage
\section{BNL - New Detector}

Mann and Murtagh \cite{BNLa} have
proposed  a long baseline experiment using the BNL-AGS which
can deliver $6 \times 10^{13}$ POT every 1.7 sec,
thus achieving $2 \times 10^{20}$ POT in 100 days.
The average neutrino energy is approximately 1 GeV and the contamination from
$\nu_e$ is about 1\%. A new neutrino beamline will
need to be constructed for this experiment.

The detectors consist of three massive imaging \c{C}erenkov
counters at 1, 3 and 20~km from the source.
These detectors will have masses of 0.8, 0.8 and 6.3 kTons respectively.
This experiment will be  a $\nu_{\mu}$
disappearance experiment using the quasi-elastic events as the signal.
The raw event rate in the far detector is 18k per 100 days of running.
The analysis will be performed by measuring the event rate in
the far, intermediate and near detectors as a function of neutrino energy.
Fig. \figbn ~contains the exclusion plots for this experiment.

\begin{figure}[hb]
\vspace{6.5in}
\label{bn}
\caption{The regions accessible to the BNL-AGS experiment with the
far detector at 10 km and 20 km.}
\end{figure}
\newpage

\section{CERN - Detector in Gran Sasso}

The CERN-SPS is an 80 to 450 GeV proton accelerator
which can conveniently be used to send a
neutrino beam from CERN to the Gran Sasso Laboratory.
The SPS-LHC transfer line is the the direction of Gran Sasso
Laboratory and the distance of 730 km makes the beam a modest
3.3 degrees below the horizontal.
The SPS accelerator is capable of delivering
$10^{20}$ POT per $10^7$ sec - year and the contamination of the
muon neutrino beam from muon  anti-neutrinos
or electron neutrinos is at the per cent level.
At 80 GeV the average energy of the neutrinos would be
approximately 6 GeV.

The ICARUS detector\cite{CERNa} in the Gran Sasso
tunnel would be a 5 kTon large
liquid Argon TPC and that would have 4k  charged current events per
$10^{20}$ POT from the CERN neutrino beam. The \mue ~analysis
would be based on $\nu_e$ appearance plus
a precise understanding of the beam contamination and the backgrounds from
$\nu_{\mu}$ neutral current interactions with $\pi^0$ faking electrons.
Whereas the results  in the \mutau ~mode would be based on a
combination of analyses;
$\nu_{\mu}$  disappearance, the neutral current to charged current ratio,
and  direct appearance of $\nu_{\tau}$.
The exclusion plots for the ICARUS/CERN experiment are given in Fig.
\figci.

\begin{figure}[hb]
\vspace{5.75in}
\label{ci}
\caption{CERN/ICARUS exclusion plots for both \mue ~and \mutau.}
\end{figure}
\newpage

The GeNIUS (GeV Neutrino-Induced Underground Shower) \cite{CERNb}
detector  is a  17 kTon (15 kTon fiducial volume)
fine-grained
sampling calorimeter to be placed, if approved,
in the Gran Sasso tunnel.
With a neutrino beam similar to the FNAL beam from
the Main Injector this detector
would have 18k charged current events for $10^{20}$ POT. The analysis
would be performed  using the neutral current to charged current ratio
for $\mutau$ and the electron-type to muon-type events for
$\mue$. The exclusion plots for this detector with
the CERN-PS producing a neutrino beam
with average energy of 5 GeV are
given in Fig. \figcg.

CERN has also discussed the possibility of sending a neutrino beam,
produced from 450 GeV protons, to SuperKamiokande, 9000~km away.
The beam would have to be aimed 44 degrees below the horizontal.
With such large separation between source and detector this experiment
will be able to study matter enhanced oscillation effects in the \mue ~mode.
An exclusion plot for this possibility can be found in Ref. \cite{CERNa}.

\begin{figure}[hb]
\vspace{6.25in}
\label{cg}
 \caption{Area of oscillation parameter space ruled out at the 90\%
confidence level for one year of running for the GeNIUS detector
and the CERN neutrino beam.}
\end{figure}

\newpage
\section{KEK - SuperKamiokande}

The possibility of sending a muon neutrino beam the 250~km  between
SuperKamiokande and
 KEK has been discuss in detail by Nishikawa\cite{KEKa}.
The KEK-PS is a 12 GeV proton accelerator which
can currently deliver $4 \times 10^{12}$
protons on target every 2.5~sec.
Therefore a modest  upgrade is  required to the KEK-PS
to deliver $10^{20}$
POT in a  period of a few years.
The average energy of the
neutrino beam is approximately 1 GeV and the contamination
of $\nu_e$ is a few percent.

SuperKamiokande is a 50 kTon water \c{C}erenkov detector which is
scheduled for completion in April of 1996.
The event rate in the 20 kTons of fiducial volume of SuperKamiokande
is 400 CC events for $10^{20}$ POT using a two radiation length target.
The analysis for the \mutau ~mode is
based on the neutral  current to charged current ratio for SuperKamiokande
and a small water \c{C}erenkov detector on the KEK site.
This requires distinguishing an EM showering
particle ({\it e} or $\gamma$) from a non-showering
particle ($\mu$ or $\pi$) which can be attained with this detector
 above a few hundred MeV,
whereas the
\mue ~mode requires distinguishing between
an electron and a $\pi^0$ which can be separated
for neutrino energies greater than 2 GeV for this detector.
The exclusion plots for this combination of accelerator and detector
are given in Fig. \figkk.

\begin{figure}[hb]
\vspace{5.5in}
\label{kk}
\caption{
The 90\% exclusion plots for the KEK-PS - SuperKamiokande experiment
for both $\nu_{\mu} \leftrightarrow \nu_{\tau}$ and
$\nu_{\mu} \leftrightarrow \nu_e $. }
\end{figure}
\section{Other and Conclusions}
Another possibility is to use the proposed new accelerator at
TRIUMF (KAON), aimed at either a new detector approximately
40 km from the source
or to SuperKamiokande in Japan. Details of these possibilities are
still under discussion. Using the SSC as a source of neutrinos has also
been discussed in conjunction with GRANDE as a detector.

Given the results from the atmospheric neutrino data,
it is important to explore the oscillation parameter region
$10^{-3}~ eV^2  <  \dmsq  <  10^{-1}~ eV^2 $
and $\s2t  > 0.4~$ \cite{other}.
Accelerator neutrinos are ideal for this purpose
because the intense beams are well understood,
with a more sharply peaked
energy spectrum and can be
manipulated as opposed to  atmospheric neutrinos.
There are a number of very exciting possibilities for experiments to
explore this region of parameter space;
let us hope that at least one of these experiments
is actually performed.
Remember that the fermion
mass question  is one of particle physics'
great mysteries beyond the Standard Model,
and any clues from  neutrinos may unleash our imagination
to further our understanding of nature.

\section{Acknowledgements}
I would like to thank R.~Bernstein, M.~Goodman, A.Mann, D.~Michael,
M.~Murtagh, K.~Nishikawa,
J.-P.~Revol and N.~Weiss for providing me with
many details for this review.
Fermilab is operated by the Universities Research
Association Inc.\ under contract with the United States Department of
Energy.

%  BIBLIOGRAPHY


\begin{thebibliography}{99}
\bibitem{other}
This region will be also explored by reactor neutrino
oscillation experiments in the \mue ~mode. See other
contributions to this workshop.
\end{thebibliography}
\end{document}